\documentclass[12pt]{article}
\usepackage{times}
\usepackage{geometry}
\usepackage[utf8]{inputenc}
\usepackage{enumitem,amssymb}
\usepackage{ragged2e}
\usepackage{setspace}
\usepackage{graphicx}
\newlist{thematic}{itemize}{8}
\setlist[thematic]{label=$\square$}
\usepackage{pifont}
\newcommand{\cmark}{\ding{51}}%
\newcommand{\done}{\rlap{$\square$}{\raisebox{2pt}{\large\hspace{1pt}\cmark}}%
\hspace{-2.5pt}}

\newcommand\apj{The Astrophysical Journal}
\newcommand\apjl{The Astrophysical Journal, Letters}     
\newcommand\aapr{Astronomy and Astrophysics Reviews}
\newcommand\pasa{Publications of the Astronomical Society of Australia}
\newcommand\solphys{Solar Physics}
\newcommand\grl{Geophys.~Res.~Lett.}

\begin{document}
\pagenumbering{gobble}
\RaggedRight
\noindent {\fontsize{16}{20} \selectfont Astro2020 Science White Paper}
\begin{center}
{\fontsize{24}{32}\selectfont Diagnostics of Space Weather Drivers Enabled by Radio Observations}
\vspace{0.5cm}

\end{center}

\vspace{0.3cm}

\normalsize

\noindent \textbf{Thematic Areas:} \hspace*{60pt} \done Planetary Systems \hspace*{10pt} $\square$ Star and Planet Formation \hspace*{20pt}\linebreak
$\square$ Formation and Evolution of Compact Objects \hspace*{31pt} $\square$ Cosmology and Fundamental Physics \linebreak
  \done Stars and Stellar Evolution \hspace*{1pt} $\square$ Resolved Stellar Populations and their Environments \hspace*{40pt} \linebreak
  $\square$    Galaxy Evolution   \hspace*{45pt} $\square$             Multi-Messenger Astronomy and Astrophysics \hspace*{65pt} \linebreak

\justifying
  
\noindent \textbf{Principal Author:} \\
Tim Bastian, National Radio Astronomy Observatory \\
Email: tbastian@nrao.edu.edu \\
Phone: (434) 296-0348 \\

\noindent \textbf{Co-authors:} \\
Hazel Bain, CIRES, University of Colorado, Boulder / NOAA SWPC \\
Bin Chen, New Jersey Institute of Techology \\
Dale E. Gary, New Jersey Institute of Technology \\
Gregory D. Fleishman, New Jersey Institute of Technology \\
Lindsay Glesener, University of Minnesota \\
Pascal Saint-Hilaire, University of California, Berkeley \\
Colin Lonsdale, MIT/Haystack
Stephen M. White, Air Force Research Laboratory \\

\noindent \textbf{\large Executive Summary}

The Sun is an active star that can have a direct impact on the Earth, its magnetosphere, and the technological infrastructure on which modern society depends. Among the phenomena that drive ``space weather'' are fast solar wind streams and co-rotating interaction regions, solar flares, coronal mass ejections, the shocks they produce, and the energetic particles they accelerate. Radio emission from these and associated phenomena offer unique diagnostic possibilities that complement those available at other wavelengths. Here, the relevant space weather drivers are briefly described, the potential role of radio observations is outlined, and the requirements of an instrument to provide them are provided: specifically, {\sl ultrabroadband imaging spectropolarimetry}. The insights provided by radio observations of space weather drivers will not only inform the science of space weather, they will pave the way for new tools for forecasting and ``nowcasting'' space weather. They will also serve as an important touchstone against which local environment of exoplanets and the impact of ``exo-space weather'' can be evaluated. 

\pagebreak
\pagenumbering{arabic}

\section{Introduction}

The term ``space weather'' refers to an array of phenomena that can disturb the interplanetary medium and/or affect the Earth and near-Earth environment. This includes recurrent structures in the solar wind (fast solar wind streams, co-rotating interaction regions), the ionizing radiation and hard particle radiations from flares, radio noise from the Sun, coronal mass ejections (CMEs), and solar energetic particles (SEPs). These drivers result in geomagnetic storms, changes in the ionosphere, and atmospheric heating which can, in turn, result in a large variety of effects that are of practical concern to our technological society: ground-level currents in pipelines and electrical power grids, disruption of civilian and military communication and navigation, spacecraft charging, enhanced atmospheric drag on spacecraft, etc. They can also disrupt the work, and endanger the lives, of humans living and working in space, on the Moon, or on Mars. The drivers of space weather are all solar in origin. An understanding of space weather phenomena and the ability to forecast such phenomena lies, in part, in gaining a fundamental understanding of these drivers.

This white paper briefly summarizes specific drivers of space weather, the potential for radio diagnostics of these drivers, and the requirements for research infrastructure investments needed to enable them. 

\section{Drivers of Space Weather and their Radio Signatures}

A number of solar phenomena have space weather impacts on the Earth and near-Earth environment. The science of space weather is focused on the fundamental physics underlying these phenomena. We touch on several to which radio observations can make unique contributions.

{\bf Solar flares}: Flares result from the catastrophic release of magnetic free energy from non-potential magnetic fields in the Sun's low corona, resulting in mass motions, plasma heating, particle acceleration, and electromagnetic radiation from radio wavelengths to hard X-rays and $\gamma$-rays. Flares produce intense electromagnetic and hard particle radiation that can have prompt effects on the ionosphere [24]. Radio emission from flares includes two key components: i) coherent radio bursts, which predominantly occur at frequencies $<3$ GHz (see below); and 2) incoherent gyrosynchrotron radiation from electrons with energies of 100s of keV to several MeV, predominantly at frequencies $>1$ GHz. As described in the white paper by {\bf Chen et al.}, coherent radio bursts enable novel diagnostics of magnetic energy release whereas observations of polarized gyrosynchrotron emission enable measurements of the magnetic field in the flaring plasma as well as the spatiotemporal evolution of the electron distribution function, thereby placing strong constraints on electron acceleration mechanisms (see also the white paper by {\bf Gary et al.}, which describes radio observations of the celebrated flare of 2017 September 10; see also Fig. 1 below).

\begin{figure}[htp]
\centering
\includegraphics[trim=0.5in 1.75in 0.5in 1.75in,clip,width=\textwidth]{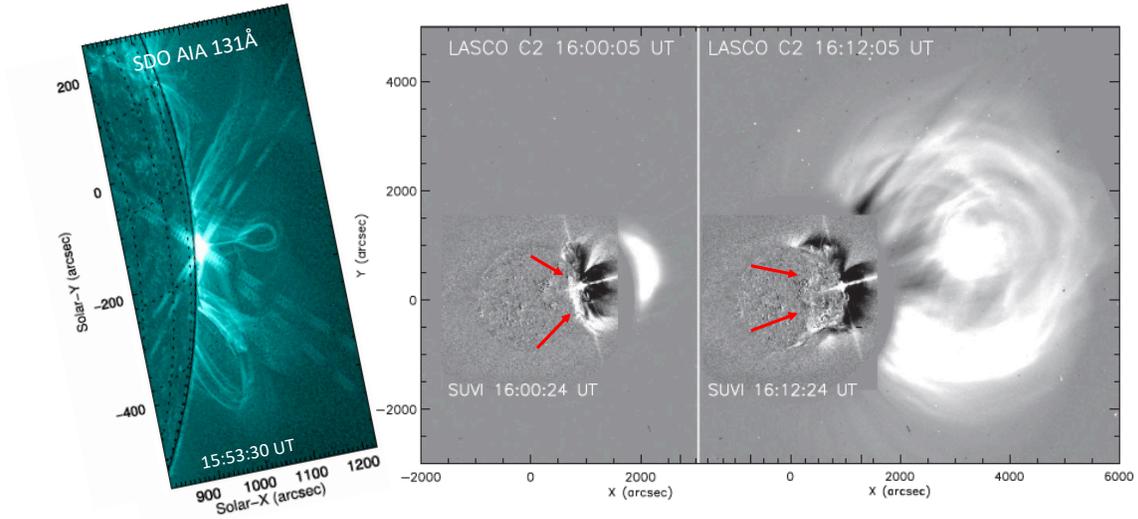}
\caption{The spectacular flare, flux rope eruption, and fast ($>3000$ km/s) CME of 2017 September 10. This X8.2 flare produced shock-driven type II radio bursts [16], a powerful SEP event [15], a long duration Fermi/LAT $\gamma-$ray event [17], and had dramatic space weather impacts on both the Earth [18] and Mars [10]. This event had it all. From [26]}
\label{fig:Sep10}
\end{figure}

{\bf Coronal Mass Ejections}: Eruptive flares and CMEs are closely intertwined. CMEs involve the destabilization and ejection of a magnetic flux rope from the low corona into the interplanetary medium [4] and may be accompanied by an erupting filament. CME masses range from $\sim 10^{14}-10^{16}$ gm and possess speeds of ~200 - 2000 km s$^-1$. The kinetic energy of a CME is therefore comparable to the energy released in large solar flares. CMEs are particularly important because they are associated with the largest geo-effective events and SEPs. With the detection of synchrotron radiation from CMEs at decimeter and meter wavelengths [2,13] a unique tool is available at radio wavelengths to detect, image, and diagnose the evolving plasma properties of CMEs, including its magnetic field. The advantages of CME detection and characterization at radio wavelengths are: i) there is no occulting disk as is the case for white light coronagraphs, so earth-directed CMEs may be detected; ii) CMEs will be detected in their nascent stages of development and can be directly associated with structures such as filament channel arcades; iii) unlike SXR and white-light observations, observations at radio wavelengths are sensitive to both thermal free-free emission from CMEs as well as nonthermal constituents. 

A long-standing goal of low frequency radio imaging arrays has been the ability to probe plasma via Faraday Rotation (FR) in the outer corona and inner heliosphere, yielding unique information on magnetic field strengths and orientations in large volumes of space.  When combined with direct or indirect tracers of electron density (i.e. white light scattering, or interplanetary scintillation measurements), FR data provide stringent constraints on the B-field properties that are not possible by any other remote sensing technique. Two key developments now offer the prospect of revolutionary measurements of FR caused by interplanetary CMEs (ICMEs). The first is the discovery of bright, diffuse linearly polarized background emission that can be assumed to be present across the entire sky at meter wavelengths [10,11]. The second is that modern low frequency imaging arrays can cover large areas of sky simultaneously, with each pixel providing a strong background linear polarization signal, typically of order 1~Jy.  This provides an extraordinarily dense spatial sampling for high precision FR measurements, capable of placing strong constraints on the magnetic fields of propagating disturbances in the solar wind, such as ICMEs. Tracking FR changes produced by ICMEs is currently a research topic, one that, when paired with density measurements using white light and/or interplanetary scintillation data, has the potential to offer wholly unique global measurements of ICME magnetic fields. 

Flares and CMEs/ICMEs both play roles in the production of SEPs. Radiation storms resulting from SEPs are a major component of space weather causing spacecraft anomalies, communications disruptions,  can pose a radiation hazard for passengers and crew on flight routes over the poles and for astronauts. However, the acceleration and transport mechanisms of SEPs are not well understood. SEP events are often associated with metric type II radio bursts, indicating CME shocks that are capable of accelerating electrons are also capable of accelerating protons to high energies [5]. Radio observations which localize sites of particle acceleration in the corona, where it is expected that particles are accelerated to the highest energies, in relation to the expansion of the CME-shock system could help distinguish between SEP acceleration mechanisms. Importantly, shock-driven type II radio burst durations [6] may also be associated with Fermi/LAT long duration $\gamma$-ray events [21]. 

{\bf Radio Bursts}: The discovery of solar radio bursts dates back to the birth of radio astronomy itself [14]. They are associated with solar active regions (type I, type III), coronal shocks (type II, type IV), and flares (type III, IV, and V). They are extremely intense and are the result of coherent emission processes, the details of which are in some cases still elusive - in large part because temporally, spectrally, and spatially resolved observations of such bursts are rare or completely absent.  The most intense of these bursts ($>10^6$ SFU, or $10^{10}$ Jy!) are not only scientifically fascinating, they can produce space weather impacts by interfering with or jamming communication and navigation technologies [1,3,7,9], particularly type IV radio continua and spectral fine structures such as solar ``spike'' bursts [25]. The detailed physics of these highly circularly polarized coherent bursts continues to be a theoretical challenge that requires polarized radio observations with appropriate spectral, spatial, and temporal resolution as well as appropriate context data. Knowing when and where they occur, and what their space weather impacts might be, is an operational challenge. 

{\bf Fast solar wind streams}: Space weather impacts are not only the result of energetic phenomena like flares, CMEs, and SEPs. Fast solar wind has its origin in coronal holes [8], regions on the Sun that are magnetically open to the interplanetary medium. Fast solar wind streams catch up with slower solar wind, resulting in a corotating interaction region (CIR) characterized by stronger magnetic fields at the fast-wind/slow-wind interface. Forward and reverse shocks may also form. CIRs and fast solar wind streams are associated with (recurrent) geomagnetic storms ([22]; see [20] for a recent review). Coronal plasma in coronal holes is cooler and less dense than surrounding corona and at centimeter and decimeter wavelengths coronal holes are dark relative to the surrounding solar atmosphere. Interestingly, for reasons that are not fully understood, coronal holes are brighter than the surrounding chromosphere at wavelengths of roughly 4 mm to 2 cm (e.g., [23] and references therein). Recent observations showing that coronal holes on the disk are bright at wavelengths longer than ~2 m, too, are also intriguing [19].  An important question is how and why the magnetic and plasma structure of a coronal hole differs from the surrounding chromosphere and corona as it extends out into the heliosphere. To make progress, imaging observations and polarimetry are needed from chromospheric heights (cm-$\lambda$) to coronal heights (m-$\lambda$).

\section{Enabling the Science of Space Weather at Radio Wavelengths}

To exploit radio observations of space weather drivers fully requires the means to perform broadband imaging spectropolarimetry from cm- to m-$\lambda$. Under quiet solar conditions, radio emission originates from the chromosphere at the shortest wavelengths and from the mid-corona at the longest wavelengths in this range. For active phenomena such as erupting flux ropes, radio CMEs, and radio bursts of type II, III, and IV, emissions may originate from considerably higher in the corona, out to several solar radii. The following high-level requirements apply:

\begin{itemize}
\item Extremely broad frequency coverage: from 50 MHz to $\sim 20$ GHz. This allows the Sun's atmosphere and the phenomena that occur there to be imaged as a system in 3D from the chromosphere well up into the corona.
\item Time resolution: the ability to observe coherent radio burst phenomena with a cadence as short as 10 ms from 50 MHz to 3 GHz and incoherent radiation with a cadence of 100 ms at higher frequencies. 
\item Spectral coverage and resolution: continuous frequency coverage is needed to measure and disentangle contributions from multiple sources and emission mechanisms. Spectral resolution of a few $\times 0.1\%$ is needed for coherent radio bursts and 1\% for incoherent emissions. 
\item Polarization: Support of full polarimetry is needed. In most cases only the Stokes I and V parameters will be needed since the Faraday depth of the corona washes out linearly polarized emission. Stokes V embodies information about the chromospheric and coronal magnetic field. 
\item Angular resolution: scattering in the Sun's corona constrains the usable angular resolution to roughly $20"/\nu_9$ where $\nu_9$ is the frequency in GHz; i.e., $1"$ at 20 GHz. 
\item Field of view: Full disk imaging at cm-$\lambda$ and imaging to several solar radii at dm- to m-$\lambda$, covering heights that are generally inaccessible to white light coronagraphs.
\item High-dynamic-range imaging: in order to exploit broadband 3D imaging of the Sun's atmosphere it must be possible to see faint phenomena such as ``EIT waves'' and radio CMEs in the presence of bright flare-related or radio-burst emission, requiring a dynamic range of order $10^4:1$.
\end{itemize}

A key point: space weather drivers cannot be considered in isolation. They are part of a complex physical system: an eruptive flare produces a fast CME. The associated flare accelerates electrons and ions to high energies, producing intense coherent radio bursts that trace magnetic reconnection and shocks, as well as incoherent gyrosynchrotron radiation that traces the accelerated electron distribution function and the flare magnetic field. The fast CME drives a strong shock through the corona into the interplanetary medium, producing bright type II radio burst that traces the shock propagation, accelerating electrons and ions to produce a gradual SEP, perhaps producing a long-duration  $\gamma$-ray afterglow in the corona of the type reported by Fermi/LAT. The drivers of these phenomena can be observed as a {\sl coupled system} using techniques offered by ultrabroadband imaging spectropolarimetry.  

These requirements are in fact met by a next-generation radioheliograph that is known as the Frequency Agile Solar Radiotelescope (FASR), a facility that has been recommended as a priority by previous decadal surveys, both the Astronomy \& Astrophysics decadals and the Solar \& Space Physics decadals. As a mid-scale-sized project, funding mechanisms have not been available to move this priority project forward. Until now. With the implementation of NSF's Mid-scale Research Infrastructure program, FASR can now be made a reality. Separate project submissions will describe the instrument in detail. 

We end by noting that comprehensive radio observations of space weather drivers will serve as a touchstone for radio observations of exoplanetary systems, and the exo-space weather they experience. With dozens of exoplanets known to exist (see, e.g., the Planetary Exoplanets Catalog\footnote{http://phl.upr.edu/projects/habitable-exoplanets-catalog}) in the so-called habitable zone (where liquid water can exist) of late-type dwarf stars, an understanding of activity on the host star and the exposure of exoplanets to the associated exo-space weather is a matter of keen interest [18].

\end{document}